\documentclass[12pt]{article}
\usepackage{epsf}

\begin{document}

\def\lesssim{\mathrel{\mathpalette\vereq<}}
\def\gtrsim{\mathrel{\mathpalette\vereq>}}
\makeatletter
\def\vereq#1#2{\lower3pt\vbox{\baselineskip1.5pt \lineskip1.5pt
\ialign{$\m@th#1\hfill##\hfil$\crcr#2\crcr\sim\crcr}}}
\makeatother

\renewcommand{\thefootnote}{\fnsymbol{footnote}}
\setcounter{footnote}{0}

\begin{titlepage}
\begin{center}

\hfill    UCB-PTH-98/20\\
\hfill    LBNL-41691\\
\hfill    hep-ph/9804291\\
\hfill    April 14, 1998\\

\vskip .5in

{\Large \bf 
Axionic Hot Dark Matter\\in the Hadronic Axion Window\footnote
{This work was supported in part by the U.S. 
Department of Energy under Contracts DE-AC03-76SF00098, in part by the 
National Science Foundation under grant PHY-95-14797.  HM was also 
supported by Alfred P. Sloan Foundation.}
}


Takeo Moroi$^{1}$ and Hitoshi Murayama$^{1,2}$

\vskip 0.05in

{\em $^{1}$Theoretical Physics Group\\
     Ernest Orlando Lawrence Berkeley National Laboratory\\
     University of California, Berkeley, California 94720}

\vskip 0.05in

{\em $^{2}$Department of Physics\\
     University of California, Berkeley, California 94720}

\vskip .5in

\end{center}

\vskip .5in

\begin{abstract}

Mixed dark matter scenario can reconcile the COBE data and the
observed large scale structure.  So far the massive neutrino with a
mass of a few eV has been the only discussed candidate for the hot
dark matter component.  We point out that the hadronic axion in the
so-called hadronic axion window, $f_{a} \sim 10^{6}$~GeV, is a perfect
candidate as hot dark matter within the mixed dark matter scenario.
The current limits on the hadronic axion are summarized.  The most
promising methods to verify the hadronic axion in this window are the
resonant absorption of almost-monochromatic solar axions from M1
transition of the thermally excited $^{57}$Fe in the Sun, and the
observation of the ``axion burst'' in water \v{C}erenkov detectors
from another supernova.

\end{abstract}
\end{titlepage}

\renewcommand{\thepage}{\arabic{page}}
\setcounter{page}{1}
\renewcommand{\thefootnote}{\arabic{footnote}}
\setcounter{footnote}{0}

\section{Introduction}

The cold dark matter (CDM) dominated universe with scale-invariant
primordial density fluctuation has been the standard theory of
structure formation.  After COBE has found the finite density
fluctuation in the cosmic microwave background radiation (CMBR), the
standard CDM scenario was found to give too much power on smaller
scales.  Many modifications to the standard CDM scenario were proposed
which solve the discrepancy: by introducing a small Hot Dark Matter
(HDM) component~\cite{aph9707285}, by ``tilting'' the primordial
density fluctuation spectrum~\cite{tilt}, by assuming a finite cosmological
constant~\cite{Lambda-CDM}, or by introducing particles
(such as $\nu_{\tau}$) whose decay changes the time of
radiation-matter equality~\cite{cdm+mnu}.  At this point, there is no
clear winner among these possibilities.\footnote{However, a large
``tilt'' is difficult to obtain in many inflationary models.
$\tau$CDM can be tested well by $B$-factory experiments in the near
future~\cite{hph9709411}.  The recent data from high-redshift 
supernovae prefer $\Lambda$CDM \cite{Saul}, but the possible evolution
of supernovae needs to be excluded by more systematic comparison 
between nearby and high-$z$ supernovae.}

In this letter, we revisit the mixed dark matter (MDM) scenario from
the particle physics point of view.  This scenario has attracted
strong interests because there has been a natural candidate for the
HDM component: massive neutrino(s).  A neutrino with a mass of a few
eV can naturally contribute to a significant fraction of the current
universe.  However, it has not been easy to incorporate the HDM
together with other neutrino ``anomalies,'' unless all three
generation neutrinos (possibly together with a sterile neutrino) are
almost degenerate, and their small mass splittings explain various
``anomalies.''  Such a scenario may be viewed as fine-tuned.
Especially, the atmospheric neutrino anomaly is quite significant
statistically now thanks to the SuperKamiokande experiment, which
suggests the mass squared difference of $\Delta m^{2} = 10^{-3} -
10^{-2}~{\rm eV}^{2}$ between the muon and tau neutrinos.  If we view
the situation from the familiar hierarchical fermion mass matrices, it
suggests the tau neutrino mass of 0.03 -- 0.1~eV, and it appears
difficult to accommodate the HDM based on massive neutrinos.

We point out that the hadronic axion~\cite{KSVZ} can be an alternative
motivated candidate for the HDM component in the MDM model.  Axion has
been proposed as a solution to the strong CP problem in the QCD, and
the hadronic axion (or KSVZ axion) is one version which predicts small
coupling of the axion to the electron.  There has been known a window
of $f_{a} \sim 10^{6}$~GeV allowed by existent astrophysical and
cosmological constraints if the axion coupling to photons is
suppressed accidentally.  This is referred to as the ``hadronic axion
window.'' Our main observation is that this window gives exactly the
right mass of $m_{a} \sim \mbox{a few eV}$ 
and the number density of the axion appropriate for the HDM
component in the MDM scenario.

\section{Hadronic Axion}

First, let us review the hadronic axion model~\cite{KSVZ}. The most
important feature of the hadronic axion is that it does not have
tree-level couplings to the ordinary quarks ($u$, $d$, $s$, $c$, $b$,
$t$) and leptons ($e$, $\nu_e$, $\mu$, $\nu_\mu$, $\tau$,
$\nu_\tau$). In this framework, we introduce new fermions which have
Peccei-Quinn (PQ) charges, while ordinary fermions do not transform
under U(1)$_{\rm PQ}$. Some of those new fermions, which we call PQ
fermions hereafter, also have SU(3)$_{\rm C}$ quantum numbers.  After
the PQ symmetry is broken spontaneously, axion $a$ appears as a
pseudo-Nambu-Goldstone boson of the PQ symmetry.

The axion $a$ couples to the photon with the operator
 \begin{eqnarray}
  {\cal L}_{a\gamma\gamma} = 
  \frac{1}{8} g_{a\gamma\gamma} a \epsilon^{\mu\nu\rho\sigma}
  F_{\mu\nu} F_{\rho\sigma}
  \equiv \frac{\alpha}{16\pi} \frac{C_{a\gamma\gamma}}{f_a}
  a \epsilon^{\mu\nu\rho\sigma} F_{\mu\nu} F_{\rho\sigma},
 \label{L_agg}
 \end{eqnarray}
 where $f_a$ is the axion decay constant. This interaction is induced
by the mixing to the light mesons ($\pi^0$, $\eta$, $\eta'$, and so
on) as well as by the triangle anomaly of the PQ fermions. By using the chiral
Lagrangian based on flavor SU(2)$_L \times$ SU(2)$_R$, we can estimate the
coefficient $C_{a\gamma\gamma}$ as~\cite{Kolb-Turner}
 \begin{eqnarray}
  C_{a\gamma\gamma}
  = \frac{E_{\rm PQ}}{N} - \frac{2(4+z)}{3(1+z)},
 \label{c_agg}
 \end{eqnarray}
where $z=m_u/m_d$ which is estimated to be 0.56 by the leading order 
perturbation in quark masses in the
chiral Lagrangian. (Hereafter, we use $z=0.56$
for our estimation, unless we discuss quantities which are sensitive
to the uncertainty in $z$.) In Eq.~(\ref{c_agg}), the first term is
from the U(1)$_{\rm em}$ anomaly of the PQ fermions, while the second
term is due to the mixing between axion and light mesons.
Simultaneously, we also obtain the formula for the axion mass as
 \begin{eqnarray}
  m_a = \frac{\sqrt{z}}{1+z} \frac{f_\pi m_\pi}{f_a}
  \simeq 6.2{\rm ~eV}\times (f_a/10^6{\rm ~GeV})^{-1},
 \label{m_a}
 \end{eqnarray}
 where $f_\pi\simeq 93{\rm ~MeV}$ is the pion decay constant, and
$m_\pi$ is the pion mass.

With this axion-photon-photon coupling, axion decays into two photons
with the lifetime
 \begin{eqnarray}
  \tau_a = \left[ \frac{\alpha^2C_{a\gamma\gamma}^2}{256\pi^3}
  \frac{m_a^3}{f_a^2} \right]^{-1}
  \simeq 1.2\times 10^{12}{\rm ~yr} 
  \times C_{a\gamma\gamma}^{-2} (m_a/10 {\rm ~eV})^{-5}.
 \label{tau_a}
 \end{eqnarray}
 Notice that the lifetime of the axion is longer than the age of the
Universe for $m_a\sim{\rm 10~eV}$ and $C_{a\gamma\gamma}\lesssim 1$,
and hence primordial axions are still in the Universe. However, as we
will see later, radiative decay of the axion may affect the background
UV photons in spite of the long lifetime.

Here, we comment that $C_{a\gamma\gamma}$ is significantly
affected by uncertainties in the chiral Lagrangian with which the
mixing effect is usually calculated.  First of all, the accuracy of the
SU(2)$_L \times$ SU(2)$_R$ chiral Lagrangian is tested up to about 5 --
10~\%. For example, by using the pion decay constant estimated from
the leptonic decay width of $\pi^\pm$, $\Gamma (\pi^0\rightarrow\gamma
+\gamma)$ is calculated to be 7.73~eV~\cite{PRL71-3629}, while
experimentally, it is measured to be $7.7\pm 0.6$~eV~\cite{PDG}. (Even
though the center value given in Ref.~\cite{PDG} is in a good
agreement, the single best measurement suggests the width to be
$7.25\pm 0.23$~eV~\cite{PLB158-81}, which is about 6~\% off from the
chiral Lagrangian prediction.) Furthermore, $f_\pi$ estimated from the
process $e^++e^-\rightarrow \pi^0+e^++e^-$~\cite{ZPC49-401} is about
10~\% smaller than the one from the leptonic decay of
$\pi^\pm$~\cite{PRL71-3629}. Therefore, we may expect 5 -- 10~\% error
in the calculation of the mixing effect from chiral Lagrangian.
Another uncertainty is from the so-called Kaplan--Manohar
ambiguity~\cite{PRL56-2004}. Within the lowest-order chiral perturbation 
theory,
$z$ is estimated to be 0.56. However, under the SU(2)$_{L}\times$ 
SU(2)$_{R}$
flavor symmetry, the quark mass matrix $M = {\rm diag}(m_u,m_d)$ and
its conjugate $(i\sigma^2)M^*(-i\sigma^2) = {\rm
diag}(m_d^*,m_u^*)$ have the same transformation properties, and hence
the following shifts are allowed: $m_u\rightarrow m_u'=m_u+\epsilon m_d^*$,
$m_d\rightarrow m_d'=m_d+\epsilon m_u^*$, where $\epsilon$ is an 
unknown parameter~\cite{PRL56-2004}.  Since the parameter
$\epsilon$ is arbitrary, $z=m_u/m_d$ cannot be determined
from the meson masses alone.\footnote{In the 
SU(3)$_{L}\times$SU(3)$_{R}$ chiral Lagrangian, the 
effect is formally higher order in quark masses, and hence $\epsilon 
\sim m_{s}/(4\pi f_{\pi})$.  Still, the ambiguity in $z$ is rather large.}
In particular, $z$ much smaller than 0.56 (or even $z=0$)
may be allowed if we take this
ambiguity into account~\cite{PRL56-2004}.  This ambiguity cannot be resolved based 
on meson masses only, but can be by using the baryon masses to some
extent.  The 
uncertainty, however, remains large~\cite{PRD52-5202}.\footnote
 {However, if $z=0$, strong CP problem is solved without introducing an
axion. Therefore, we do not consider this possibility in this letter.}
 The mixing contribution to $C_{a\gamma\gamma}$ is affected by this
uncertainty in $z$.


As we will see later, $C_{a\gamma\gamma}$ is constrained to be less
than 0.01 -- 0.1 from astrophysical arguments for the axion decay
constant we are interested in. In general, $C_{a\gamma\gamma}\ll 1$ is
possible if we adopt an accidental cancellation. With the lowest order
chiral Lagrangian, cancellation occurs when $E_{\rm
PQ}/N=2(4+z)/3(1+z)\simeq 1.95$, but this estimation may not be so
reliable. We believe that a better understanding of the quark masses
is necessary to pin down the value of $E_{\rm PQ}/N$ for
the accidental cancellation. With the current best knowledge, it is 
clear
that the cancellation is quite possible for models with $E_{\rm
PQ}/N\sim 2$ if we take the effects we discussed above into account. In
particular, the possibility of the value obtained in grand-unified 
theories ($E_{\rm PQ}/N=8/3$) may not be excluded.

The axion is also coupled to fermions: ${\cal L}_{aff}=g_{aff}
a\bar{f}i\gamma_5f$, which can again be estimated by using the chiral
Lagrangian. Importantly, the hadronic axion does not couple to
ordinary quarks and leptons at the tree level. Therefore, in
particular, the axion-electron-electron coupling has an extra loop
suppression factor~\cite{PLB316-51}:
 \begin{eqnarray}
  g_{aee} = \frac{3\alpha^2}{4\pi^2} \frac{m_e}{f_a}
  \left\{ \frac{E_{\rm PQ}}{N} \ln (f_a/m_e) 
  - \frac{2(4+z)}{3(1+z)} \ln (\Lambda_{\rm QCD}/m_e) 
  \right\}.
 \label{g_aee}
 \end{eqnarray}
 On the other hand, mixing effects induce an axion-nucleon-nucleon
coupling, even though the axion-quark-quark coupling vanishes at the tree
level for a hadronic axion:
 \begin{eqnarray}
  g_{aNN} = \frac{m_N}{f_a}
  \left\{ ( F_{A0} \mp F_{A3} ) \frac{1}{2(1+z)} + 
  ( F_{A0} \pm F_{A3} ) \frac{z}{2(1+z)} \right\},
 \end{eqnarray}
 where $m_N\simeq 940{\rm ~MeV}$ is the nucleon mass, and upper
(lower) sign is for neutron (proton).
The axial-vector isovector contribution has been quite well
understood to be $F_{A3}\simeq -1.25$ from the neutron
$\beta$-decay. Isoscalar part $F_{A0}$ used to 
be more ambiguous, since this quantity depends on the flavor-singlet
axial-vector matrix element $S$ (with $S\equiv\Delta u+\Delta d+\Delta
s$ in Ref.~\cite{hph9601280}) as $F_{A0}\simeq -0.67S-0.20$, where the
constant piece is determined by the hyperon $\beta$-decay.  In
Ref.~\cite{hph9601280}, however, $S$ was estimated from experimental
data including higher order QCD corrections, resulting in $S=0.27\pm
0.04$. Even though possible systematic uncertainties are not included
in this calculation, we use this result as a reference when we discuss
axion-nucleon-nucleon coupling.
 Because of these interactions, $f_a$ is constrained by the axion
emission from SN1987A.

\section{Constraints on Hadronic Axion}


Next, we summarize the constraints on the hadronic axion. In the later
discussion, we will be interested in the case of $f_a\sim 10^{6}{\rm
~GeV}$ so that hadronic axion becomes a good candidate of the
HDM. Therefore, in this section, we pay an attention to this case.

Most importantly, the coupling of the hadronic axion to the electron
is loop suppressed, as can be seen in Eq.~(\ref{g_aee}).  Therefore,
the constraint on the axion-electron-electron coupling from the
cooling of the red giant~\cite{redgiant,PRD51-1495} can be evaded.
One can compare the current best upper limit ($g_{aee}\lesssim
2.5\times 10^{-13}$~\cite{PRD51-1495}) with Eq.~(\ref{g_aee}), and see
that $g_{aee}$ for $f_a\sim 10^{6}{\rm ~GeV}$ is smaller than the
bound from the red giant for values of $E_{\rm PQ}/N \lesssim 7$.

A non-trivial constraint comes from the emission of the axion from a
supernova. If an axion couples to nucleons strongly, the axion can be
produced in the core of the supernova, and the axion emission may
affect the cooling process of the supernova. In particular, the
Kamiokande group and the IMB group measured the flux and duration time
of the neutrino burst emitted from the SN1987A, and their results are
consistent with the generally accepted theory of the core
collapse. Therefore, they confirmed the idea that most of the energy
released in the cooling process is carried off by neutrinos. If axion
carries away too much energy from the supernova, it would conflict
with those observations. The axion flux from the supernova can be
suppressed enough in two parameter regions. If axion-nucleon-nucleon
interaction is weak enough, the axion cannot be effectively produced
in the core of the supernova.  Quantitatively, for $f_a\gtrsim
10^9$~GeV, the axion flux can be small enough not to affect the
cooling process~\cite{SN1987A}. On the contrary, if the axion
interacts strongly enough, the mean free path of the axion becomes
much shorter than the size of the core, and hence the axions cannot
escape from the supernova. In this case, axion is trapped inside the
so-called ``axion sphere,'' and the axion emission is also
suppressed. (In this case, axions are emitted only from the surface of
the axion sphere; this type of the axion emission is often called
``axion burst.'')  Quantitatively, for $f_a\lesssim 2\times 10^6$~GeV
(or equivalently, $m_a\gtrsim 3$~eV), the axion luminosity from
SN1987A is suppressed enough~\cite{SN1987A}.

For $f_a\lesssim 2\times 10^6$~GeV suggested from the cooling of
supernova, we have another constraint from the detection of
axions in water \v{C}erenkov detectors. In this parameter region,
axion flux from the axion burst is quite sizable for its detection,
even though it does not affect the cooling of SN1987A.  If the
axion-nucleon-nucleon coupling is strong enough, axions may excite the
oxygen nuclei in the water \v{C}erenkov detectors (${\rm
^{16}O}+a\rightarrow {\rm ^{16}O^*}$), followed by radiative decay(s)
of the excited state. If this process had happened, the Kamiokande 
detector
should have observed the photon(s) emitted from the decay of ${\rm
^{16}O^*}$.  Due to the non-observation of this signal, $f_a\lesssim
3\times 10^5$~GeV is excluded~\cite{PRL65-960}.

Another class of constraint is from the axion-photon-photon coupling.
Because of this coupling, axion can be produced in Primakoff process
in the presence of external electromagnetic field, and it also decays
into two photons, which result in constraints on the (model-dependent)
axion-photon-photon coupling.

One of the important constraints comes from the cooling of the
horizontal-branch (HB) stars. If the axion-photon-photon coupling is
too strong, axions are produced in the HB stars through the Primakoff
process, and the emission of the axions affects the cooling of the HB
stars. Then, the lifetime of the HB stars becomes shorter than the
standard prediction, and the number of the HB stars are suppressed.
However, the number of the HB stars are in a good agreement with
theoretical expectations, and hence we obtain the upper bound on the
axion-photon-photon coupling~\cite{HBstar}:
 \begin{eqnarray}
  g_{a\gamma\gamma} \lesssim 6\times 10^{-11} {\rm ~GeV}^{-1}.
 \label{HB_gagg}
 \end{eqnarray}
 The important point is that $g_{a\gamma\gamma}$ has two sources: the
electroweak anomaly of the PQ fermions and the mixing between the
axion and light mesons (see Eqs.~(\ref{L_agg}) and (\ref{c_agg})).
Furthermore, the mixing effect is usually calculated by using the
chiral Lagrangian, and there is some uncertainty as discussed earlier.
Therefore, it is difficult to convert the constraint (\ref{HB_gagg})
to the constraint on the PQ scale $f_a$. In fact, due to the model
dependence, we only have an upper bound on the coefficient
$C_{a\gamma\gamma}$:
 \begin{eqnarray}
  C_{a\gamma\gamma} \lesssim 0.05 \times (f_a/10^6{\rm ~GeV}).
 \end{eqnarray}
 Notice that, in principle, any value of $f_a$ can be viable with the
cooling of the HB stars, if we adopt an accidental cancelation in
$C_{a\gamma\gamma}$.

Another important constraint is from the effects of the radiative
decay of the axion on the background UV photons. As noted in
Eq.~(\ref{L_agg}), axion is coupled to photons, and it decays into two
photons with the lifetime given in Eq.~(\ref{tau_a}). Even though the
lifetime is longer than the age of the Universe, some fraction of the
axion decays and we may see the emission line.

Constraint from the UV extragalactic light is discussed in
Refs.~\cite{PRL59-2489,PRD44-3001, APJ414-449}. Since the lifetime of
the axion is longer than the age of the Universe, intensity of the
photon is proportional to the inverse of the lifetime. Therefore, the
intensity becomes smaller as the axion-photon-photon coupling gets
weaker, and non-observation of the signal sets an upper bound on
$C_{a\gamma\gamma}$. Overduin and Wesson looked for the emitted photon
from the axion in the extra galactic light, and no signal of the axion
was found. From their observation, they derived the upper bound on
$C_{a\gamma\gamma}$ of $0.72$ ($m_a=$ 3.8~eV) to $0.014$ ($m_a=$
13.0~eV)~\cite{APJ414-449}.

More stringent constraint may be obtained if we observe the photons
emitted from the axions in clusters of galaxies.
At the center of a cluster, axions are expected to be gravitationally
trapped, and its density is more enhanced than the cosmological 
density. Therefore, the emission lines
may be more intense than the one from the extra galactic sources, and
the constraint may be more stringent. With three samples of clusters,
Ressell obtained the upper bound on $C_{a\gamma\gamma}$ of $0.12$
($m_a=$ 3.5~eV) to $0.008$ ($m_a=$ 7.5~eV)~\cite{PRD44-3001}, which is
about one order of magnitude more stringent than the constraint from
extra galactic background light. However, it is possible that the
lines of sight of the particular galactic clusters are obscured by
absorbing material, resulting in too stringent
constraint~\cite{APJ414-449}.  If we adopt this argument, this
constraint may be evaded.

 \begin{figure}[t]
 \centerline{\epsfxsize=0.6\textwidth\epsfbox{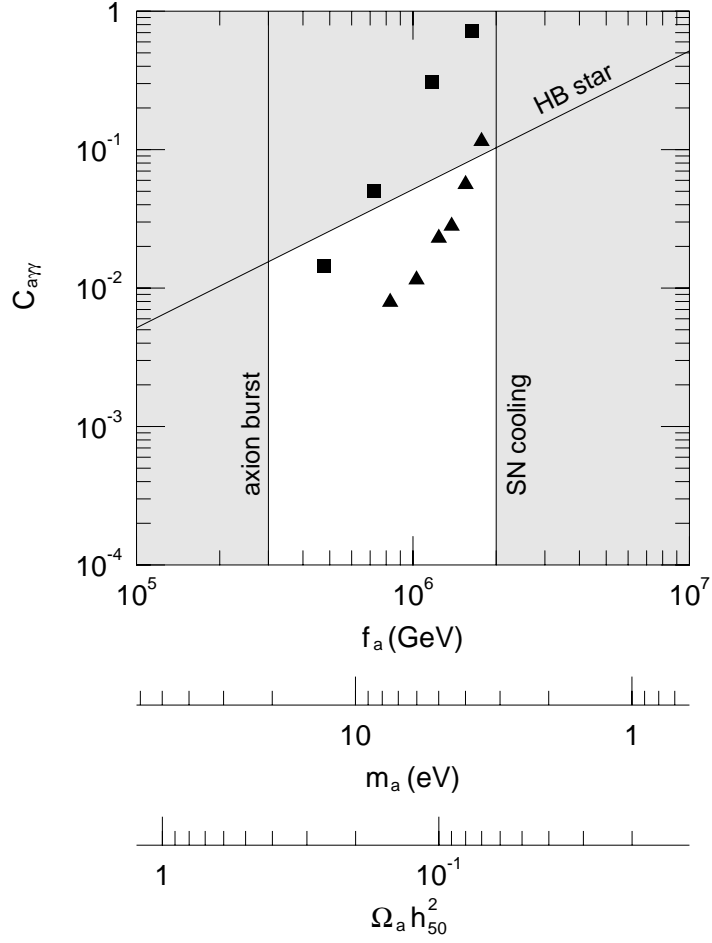}}
 \caption{Astrophysical constraints on the hadronic axion model from
the cooling of the supernova, axion burst, cooling of the HB stars,
extragalactic light~\protect\cite{APJ414-449} (square), and emission
line in clusters of galaxies~\protect\cite{PRD44-3001} (triangle).
Shaded region is excluded, and $C_{a\gamma\gamma}$ larger than squares
and triangles are inconsistent with observations for fixed value of
$f_a$.}
 \label{fig:cagg}
 \end{figure}

All the constraints mentioned above are summarized in
Fig.~\ref{fig:cagg}. As we have discussed, the hadronic axion with the
axion decay constant in the following range is still viable with all 
the
astrophysical constraints (if $C_{a\gamma\gamma}$ is small enough):
 \begin{eqnarray}
  3\times 10^5 {\rm ~GeV} \lesssim f_a \lesssim 
  2\times 10^6 {\rm ~GeV}
  ~~~ (20 {\rm ~eV} \gtrsim m_a \gtrsim 3 {\rm ~eV}).
 \label{fa-window}
 \end{eqnarray}
 Notice that the constraints based on the axion emission from SN1987A
is relatively model-independent. That is, in the hadronic axion model,
the axion-nucleon-nucleon coupling is from the mixing between the
axion and the light mesons, and hence it is independent of the
U(1)$_{\rm PQ}$ charges of the PQ fermions.\footnote{It does suffer
from the uncertainty in $z$ mentioned earlier, however
\cite{PLB316-51}.}

Finally, we comment on the constraint from the cooling of the red
giants and the HB stars due to the axion-nucleon-nucleon
coupling~\cite{PRL66-2557}.  The axion-nucleon-nucleon coupling would
allow an axion emission from red giants and the HB stars, and cause an
additional energy loss rate which is proportional to $m_a^2$. This
extra energy loss changes the brightness of these stars, and it also
modifies the relative numbers of the red giants to the HB
stars. Observed values of these quantities are in reasonable
agreements with theoretical calculations, and hence we can obtain the
upper bound on the axion emission rate.  The constraint is quite
sensitive to the flavor-singlet axial-vector matrix element
$S\equiv\Delta u+\Delta d+\Delta s$, since axion-nucleon-nucleon
coupling depends on $S$. For $S=0.27$ as suggested in
Ref.~\cite{hph9601280}, axion mass smaller than about 12~eV is still
allowed,\footnote{The authors of Ref.~\cite{PRL66-2557} used $F_{A0} =
  -0.67 S - 0.23$ from hyperon $\beta$ decay without SU(3) breaking
  effects.  A direct measurement, however, suggests $-0.67 S - 0.20$
  \cite{Hseuh}, and makes the $S$ in their plot effectively smaller by
  0.04.}  and larger axion mass is still viable if we adopt sizable
uncertainty in $S$~\cite{PRL66-2557}.  Therefore, we concluded that
most of the parameter region for the axionic HDM is still alive.

\section{Thermal Relic of Hadronic Axion}

We have seen in the previous section that the hadronic axion with the
decay constant in the window $3\times 10^5 {\rm ~GeV} \lesssim f_a
\lesssim 2\times 10^6 {\rm ~GeV}$ is astrophysically allowed as long as
the axion-photon-photon coupling is sufficiently small.  Now, we are
in the position to discuss how the hadronic axion can be a good
candidate for HDM. For this purpose, remember that the relevant mass
range for the HDM is 1~eV -- 10~eV, corresponding to the PQ scale of
$f_a\sim 10^{6}{\rm ~GeV}$ (see Eq.~(\ref{m_a})), if the axion
decouples around the same stage as when the neutrinos do.

For $f_a\sim 10^{6}{\rm ~GeV}$, the most important source of the
primordial axions is the thermal production, rather than the coherent
oscillation~\cite{PRL59-2489,PLB316-51}. Because of the couplings to
nucleons (and to pions), axion are thermalized when $T\gtrsim
30-50{\rm ~MeV}$ for $f_a\sim 10^{6}{\rm ~GeV}$.  In the most recent
calculation~\cite{PLB316-51}, the axion density is estimated as
$[\rho_a/\rho_\nu]_{T\sim {\rm 1~MeV}}\simeq 0.4 - 0.5$, with $\rho_a$
($\rho_\nu$) being the energy density of the axion (neutrino of one
species), or equivalently,
 \begin{eqnarray}
  \frac{n_a}{s} \simeq 0.02,
 \label{Yield_a}
 \end{eqnarray}
 where $n_a$ is the number density of axion, and $s$ is the total
entropy density. (Here, we used $[\rho_a/\rho_\nu]_{T\sim {\rm
1~MeV}}=0.45$.) Then, the relic density of the axion is given by
 \begin{eqnarray}
  \Omega_a = \frac{m_an_a}{\rho_c}
  \simeq 0.2 \times h_{50}^{-2} (m_a/{\rm 10~eV}),
 \label{Omega_a}
 \end{eqnarray}
 where $h_{50}$ is the Hubble constant in units of 50~km/sec/Mpc.
Thus, for $m_a\sim 10 {\rm ~eV}$, $\Omega_a$ can be 0.1 -- 0.2 which
is the requirement for the HDM in the MDM scenario.  Note that the
hadronic axion discussed here is a thermal relic with its mass of
$\sim 10{\rm ~eV}$.  Therefore, the axion here is a relativistic
particle when the galactic scale crossed the horizon, and behaves as
HDM.\footnote{It is interesting to note that the axion decay constant
required in this scenario is rather close to the so-called messenger
scale in models with gauge mediation of supersymmetry breaking
\cite{DNS}, as well as the mass scale of the right-handed neutrino in
the sneutrino CDM scenario \cite{HMM}.  It is conceivable that the
field $S$ which generates the supersymmetric and
supersymmetry-breaking masses of the messengers carry the PQ charge
and the messengers are the PQ fermions.  The same field $S$ can
generate the required size of the right-handed neutrino mass in the
sneutrino CDM scenario.  The original scale of supersymmetry breaking,
however, needs to be raised to make the gravitino heavier than the
sneutrino, which can be achieved by making the messenger U(1) coupling
constant somewhat small, $\sim 0.03$.}
 
Comparing with Eq.~(\ref{Omega_a}), the window (\ref{fa-window})
is exactly where the axion has
the right mass and number density to be the HDM component in the 
MDM scenario. 

One may worry about the effect of the hadronic axion on the big-bang
nucleosynthesis (BBN). At the time of the BBN, energy density of the
axion is sizable ($[\rho_a/\rho_\nu]_{T\sim {\rm 1~MeV}}\simeq 0.4 -
0.5$), and it raises the freeze out temperature of the neutron by
speeding up the expansion rate of the Universe. As a result, in our
case, more $^4$He is synthesized than in the standard BBN
case~\cite{PLB316-51}.  A few years ago, the observed value of the
primordial $^4$He abundance seemed to be unacceptably smaller than the
theoretical prediction~\cite{BBNcrisis}. If this was true, a hadronic
axion with $f_a\sim 10^{6}{\rm ~GeV}$ could be extremely disfavored.
However, the current situation is more controversial. Recently, both
for D and $^4$He, several new measurements have been done to determine
their primordial abundances, but the results are not consistent with
each other; some group reports low D abundance while the other results
are much higher, and the same for $^4$He. In particular, if we adopt a
high value of the observed $^4$He abundance~\cite{APJS108-1}, our
scenario is consistent with the BBN. Since it is too premature to
judge which measurements are reliable, we do not expect any solid
argument based on the BBN which rules out the hadronic axion as the
HDM component in the MDM scenario.

\section{Prospect for Detecting Hadronic Axion}

So far, we have seen that the hadronic axion in the current allowed
parameter range almost automatically becomes appropriate for HDM.  As
discussed, this scenario is consistent with all the astrophysical
constraints, if the axion-photon-photon coupling is suppressed enough,
presumably by an accidental cancellation.

However, this scenario can be tested in the future in several
observations. One possibility is to use the observation of the diffuse background
UV photon. Accuracy of the current observation just excluded the
axion-photon-photon coupling down to $C_{a\gamma\gamma}\lesssim
0.1-0.01$, as we have discussed. However, if the background photon
spectrum will be well measured with a better resolution, the emission
line from the axion decay may be found in the background photon
spectrum.  However, as we emphasized, $C_{a\gamma\gamma}$ is a
model-dependent parameter.  Therefore, a non-observation of the signal
cannot exclude the possibility of hadronic axion HDM definitively,
because of a possible accidental cancellation in $C_{a\gamma\gamma}$.

Therefore, a detection of hadronic axion which does not rely on 
axion-photon-photon coupling is strongly favored.
One such possibility is to detect an axion burst from a
future supernova at SuperKamiokande (or, in general, water
\v{C}erenkov detectors). An important point is that newer water
\v{C}erenkov detectors (like SuperKamiokande) have much larger
fiducial volume than Kamiokande, and hence we can expect a larger
event rate. Therefore, a hadronic axion with $f_a\sim 10^6$~GeV can be
tested with a future supernova of the size and the distance of
SN1987A, even though SN1987A could not exclude this possibility.

Calculation of the event rate suffers from the uncertainties in
the axion-nucleon scattering cross section and modeling of supernovae.
However, the detection of the signal appears plausible. For example,
by rescaling the result given in Ref.~\cite{PRL65-960}, we expect a
few events at SuperKamiokande for a supernova of the same size as
SN1987A for $f_a\sim 10^6$~GeV. Of course, if a new supernova will be
closer than SN1987A, we can expect larger number of events, and
the hadronic axion HDM can be tested much easier.

Another interesting novel idea is due to Moriyama~\cite{Moriyama}.  
In the Sun, thermally excited $^{57}$Fe nuclei can decay by 
emitting axions.  Thanks to the Doppler broadening of the axion energy 
due to the thermal motion of $^{57}$Fe, the same nuclide can 
resonantly absorb the axion.  The detection rate was estimated and can 
be as high as 1~event/day/kg or more.  A search was already 
performed along this line \cite{Krcmar} even though they used a small 
target of 0.03~g to detect 14.4~keV gamma-ray escaping the target 
rather than the bolometric method suggested.  They obtained an upper bound 
on the axion mass of 745~eV.  Another experimental effort to detect 
solar axions is underway and may reach the axion mass as small as 3~eV 
in a few years \cite{Moriyama2}.

\section{Conclusions}

In this letter, we have pointed out that the hadronic axion in the 
hadronic axion window ($f_a\sim 10^6{\rm ~GeV}$) can automatically be 
a good candidate of the Hot Dark Matter component in the mixed dark 
matter scenario.  In order to evade an astrophysical constraint from the 
background UV light, axion-photon-photon coupling has to be suppressed 
in the hadronic axion window, probably by an accidental cancellation.  
This scenario may be tested by detecting the axion burst from a
future supernova in water \v{C}erenkov detectors, or detecting solar 
axions using resonant absorption.

\section*{Acknowledgements} 

We thank Shigetaka Moriyama for letting us know his work.
HM thanks Gia Dvali, Markus Luty, Joe Silk, and Mahiko Suzuki for
useful conversations.  This work was supported in part by the
U.S. Department of Energy under Contracts DE-AC03-76SF00098, in part
by the National Science Foundation under grant PHY-95-14797.  HM was
also supported by Alfred P. Sloan Foundation.

\end{document}